# PD-ADSV: An Automated Diagnosing System Using Voice Signals and Hard Voting Ensemble Method for Parkinson's Disease

Paria Ghaheri[1], Ahmadreza Shateri[1], Hamid Nasiri[2,*]

[1] Electrical and Computer Engineering Department, Semnan University, Semnan, Iran
[2] Department of Computer Engineering, Amirkabir University of Technology (Tehran Polytechnic), Tehran, Iran.

Correspondence should be addressed to Hamid Nasiri; h.nasiri@aut.ac.ir

**Abstract**

Parkinson's disease (PD) is the most widespread movement condition and the second most common neurodegenerative disorder, following Alzheimer's. Movement symptoms and imaging techniques are the most popular ways to diagnose this disease. However, they are not accurate and fast and may only be accessible to a few people. This study provides an autonomous system, i.e., PD-ADSV, for diagnosing PD based on voice signals, which uses four machine learning classifiers and the hard voting ensemble method to achieve the highest accuracy. PD-ADSV is developed using Python and the Gradio web framework.

**Code metadata**

| Nr | Code metadata description | |
|---|---|---|
| C1 | Current code version | *V1.0* |
| C2 | Permanent link to code/repository used for this code version | *https://github.com/Ahmadreza-Shateri/PD_ADSV* |
| C3 | Permanent link to reproducible capsule | *https://codeocean.com/capsule/8141825/tree/v1* |
| C4 | Legal code license | *GNU General Public License v3.0* |
| C5 | Code versioning system used | *git* |
| C6 | Software code languages used | *Python* |
| C7 | Compilation requirements, operating environments and dependencies | *Python 3.8 or later*<br>*Keras, TensorFlow, Pandas, NumPy, Gradio, Scikit-learn, XGboost, Lightgbm, Altair* |
| C8 | If available, link to developer documentation/manual | - |
| C9 | Support email for questions | h.nasiri@aut.ac.ir |

**Software metadata**

| | |
|---|---|
| Current software version | *1.5.3* |
| Permanent link to executables of this version | *https://github.com/Ahmadreza-Shateri/PD_ADSV* |
| Legal Software License | *GNU General Public License v3.0* |
| Operating System | *Microsoft Windows 7 (or later)*<br>*Mac OS 10.12.6 (Sierra or later)*<br>*Linux* |
| Installation requirements & dependencies | *4 GB of memory*<br>*1 GB of free disk space* |

## 1. Introduction

Parkinson's disease (PD) is the second most prevalent neurodegenerative disorder after Alzheimer's and a leading cause of neurological morbidity worldwide [1], [2]. In most cases, Parkinson's disease can be diagnosed based on the patient's motor symptoms [3] or through alternative neuroimaging methods such as PET scans and MRI [4]; However, in addition to being costly, time-consuming, and inaccessible to the general public, these procedures are not remarkably accurate when diagnosing patients. Recent studies indicate that nearly 90 percent of PD patients suffer from vocal disorders as one of its first symptoms [5]. Voice and speech issues are characterized by decreased absolute speech volume and pitch variation, breathiness, tremor, hoarse voice quality (roughness), variable speech rates, and imprecise articulation [6]. Therefore, analyzing the voice signals of Parkinson's patients is a vital step in the early diagnosis of this disorder.

According to previous studies [7]–[9], Replicated Acoustic Features of the voice signals of Parkinson's disease patients have been shown to provide crucial and valuable information for diagnosing PD. Consequently, these features were implemented into the software introduced in this paper. PD-ADSV employs the method proposed by Ghaheri et al. [9], which classified extracted features using Extreme Gradient Boosting (XGBoost), Light Gradient Boosting Machine (LightGBM), Gradient Boosting, and Bagging. Furthermore, the Hard Voting Ensemble method was used based on the performance of the four classifiers.

## 2. Gradient Boosting Decision Tree

An ensemble of weak learners, primarily Decision Trees, is utilized in Gradient boosting to increase the performance of a machine learning model [10]. The Gradient boosting decision tree (GBDT) technique enhances classification and regression tree models using gradient boosting. Data scientists frequently employ GBDT to achieve state-of-the-art results in various machine learning challenges [11].

## 3. Extreme Gradient Boosting

Extreme Gradient Boosting (XGBoost) is an improved gradient tree boosting system presented by Chen and Guestrin [12] featuring algorithmic advances (such as approximate greedy search and parallel learning [13], [14]) and hyper-parameters to enhance learning and control overfitting [15], [16]. In recent years, XGBoost has been widely utilized by researchers, and its performance in a range of Machine Learning (ML) challenges has been remarkable [17]–[24].

## 4. LightGBM

Researchers from Microsoft and Peking University initially developed the LightGBM [25] to address the efficiency and scalability issues with GBDT (Gradient Boosting Decision Tree) and XGBoost when applied to problems with high-dimensional input features and large datasets [26]. The LightGBM algorithm incorporates two innovative strategies: gradient-based one-side sampling (GOSS) [27] and exclusive feature bundling (EFB) [28], [29].

## 5. Bagging

Leo Breiman [30] proposed bagging in 1994 as a resampling technique for driving single classifiers using bootstrap samples. Bagging reduces variance and overfitting, improves ML algorithm accuracy and consistency, and preserves DT bias [31].

## 6. Software features

PD-ADSV is built on four Machine Learning classifiers: XGBoost, LightGBM, Gradient Boosting, and Bagging. The Hard Voting Ensemble Method has also been used to achieve the highest accuracy using patients' voice signals. This software implements machine learning algorithms utilizing Python and the Gardio web-based visual interface, providing maximum performance and user-friendliness [32]. The developed software uses Python version 3.9.7 and Gradio framework version 3.11.0.

To train the models, a dataset of replicated acoustic features of the voice signals [7] was collected as follows: Maintaining a steady phonation of the /a/ vowel at a comfortable pitch and volume is the vocal task. This phonation must be held for a minimum of five seconds for every breath. Each individual repeats the exercise three times, and each repetition is considered a replication. The voice data is recorded using a portable computer equipped with an external sound card (TASCAM US322) and a cardioid-pattern headband microphone (AKG 520). The Audacity software (release 2.0.5) makes a digital recording with a sampling rate of 44.1 kHz and a resolution of 16 bits/sample. 32 Acoustic features are extracted from the voice signals: 5 Harmonic-to-noise-ratio (HNR), 13 Derivatives of Mel frequency cepstral coefficients (Delta), 13 Mel frequency cepstral coefficients (MFCC), and Glottal-to-Noise Excitation Ratio (GNE).

In general, this software has two steps: 1) receiving the user's voice signals; 2) performing classification, i.e., detecting whether the person has Parkinson's disease signs or not. In the first step, the user uploads the voice signal sample (Fig. 1). Then, in the second step, the input is classified by four trained classifiers, including XGBoost, LightGBM, Gradient Boosting, and Bagging. Moreover, to utilize the advantageous characteristics of each classifier to enhance accuracy, the weighting was set depending on each classifier's performance. Finally, Hard Voting Ensemble Method determined the final prediction (Fig. 2). According to [9], the model utilized in PD-ADSV based on "Parkinson Dataset with Replicated Acoustic Features" [7] achieved an accuracy of 85.42%.

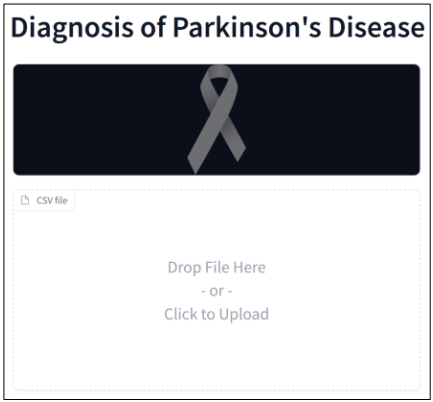
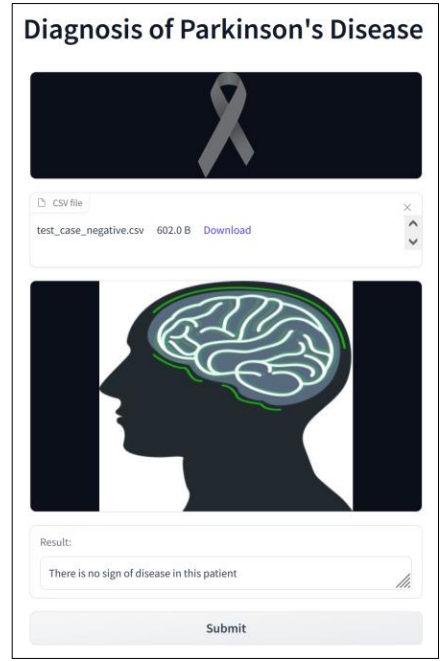

**Fig. 1.** Home page of PD-ADSV.   **Fig. 2.** The result of PD-ADSV.

## 7. Impact overview

As previously mentioned, this software can diagnose Parkinson's disease using voice signals. According to the simple user interface, it is accessible to all social classes. Anyone can record their voice and use this program to determine whether or not they have signs of PD. As depicted in Fig. 1, the user uploads their dataset of voice signals, and the result is displayed in less than 0.7 seconds.

Motor symptoms and imaging tests, such as MRI, brain ultrasonography, and PET scans, are usually used to diagnose this disease [33]. However, in addition to their low accuracy and high prices, these techniques are prohibitively hard to perform and are not easily accessible to the general public. In contrast, PD-ADSV, with its remarkable speed and accuracy, can significantly assist healthcare providers, particularly neurologists, in detecting Parkinson's disease.

As mentioned, recording the voices of PD patients for this program does not require special equipment and may be readily offered to patients in any hospital or medical center. Consequently, patients record their voices using the equipment described previously. After classifying the extracted acoustic features of voice signals, PD-ADSV determines whether or not the individual has PD signs.

## 8. Conclusion

This article introduces PD-ADSV, an automated diagnosing system that utilizes voice signals to detect PD. According to its user-friendliness, high accuracy, and availability to everyone, this software can be used in any healthcare center and greatly help doctors. It is implemented based on machine learning methods and reached an accuracy of 85.42% using "Parkinson Dataset with Replicated Acoustic Features."

## Declaration of competing interest

The authors declare that they have no known competing financial interests or personal relationships that could have appeared to influence the work reported in this paper.

## CRediT authorship contribution statement

**Paria Ghaheri:** Conceptualization, Methodology, Software, Investigation, Writing - Original Draft. **Ahmadreza Shateri:** Conceptualization, Methodology, Software, Validation, Investigation, Visualization. **Hamid Nasiri:** Conceptualization, Methodology, Validation, Writing - Review & Editing, Supervision.